\documentstyle[12pt,epsf]{article}

\begin{document}

\title{The length of a typical Huffman codeword\thanks{Submitted
 to IEEE Transactions on Information Theory.}}

\author{R\"udiger Schack\thanks{Supported by
         a fellowship from the Deutsche Forschungsgemeinschaft.}\\
   Department of Physics and Astronomy,
   University of New Mexico\\
   Albuquerque, NM 87131--1156}

\date{January 11, 1993}

\maketitle

\begin{abstract}
If $p_i$ ($i=1,\dots,N$) is the probability of the $i$-th letter of a
memoryless source, the length $l_i$ of the corresponding binary
Huffman codeword can be very different from the value $-\log p_i$.
For a {\it typical\/} letter, however, $l_i\approx-\log p_i$.  More
precisely, ${\displaystyle P_m^-=\sum_{j\in\{i|l_i<-\log
p_i-m\}}p_j<2^{-m}}$ and ${\displaystyle P_m^+=\sum_{j\in\{i|l_i>-\log
p_i+m\}}p_j<2^{-c(m-2)+2}}$ where $c\approx2.27$.
\end{abstract}

\section*{Introduction}

Concepts from information theory gained new importance in physics
\cite{Zurek1989,Schack1992} when Bennett \cite{Bennett1982} realized
that Landauer's principle \cite{Landauer1961}, which specifies the
unavoidable energy cost $k_BT\ln2$ for the erasure of a bit of
information, is the clue to the solution of the problem posed by
Maxwell's demon. This problem can be summarized as follows: A demon
knows initially that a system is in the $i$-th possible state
($i=1,\dots,N$) with probability $p_i$. The demon then finds the
actual state state of the system---thereby lowering the system's
entropy by the amount $H=-\sum p_i\log p_i$. This is in apparent
violation of the second law of thermodynamics, since the entropy
decrease corresponds to a
free-energy increase $\Delta F=Hk_BT\ln2$ that can be
extracted as work. Bennett solved this inconsistency by noting that in
order to return to its original configuration the demon must erase
its record of the system state. The second law is saved since, due to
Shannon's noiseless coding theorem, the average length of the demon's
record cannot be smaller than $H$. Therefore, the Landauer erasure
cost cancels the extracted work on the average.

If the demon wants to operate with maximum efficiency, it must use an
optimal coding procedure, i.~e., Huffman coding \cite{Huffman1952}.
In this context, the question arises as to how the record length $l_i$
for the $i$-th state can be interpreted. Zurek \cite{Zurek1989}
discusses two alternative (sub-optimal) coding procedures for the
demon: minimal programs for a universal computer, where the record
length is the algorithmic complexity \cite{Chaitin1987} of the state;
and Shannon-Fano coding, where the record length is determined by the
state's probability through the inequality $-\log p_i\leq l_i<-\log
p_i+1$. The length of a Huffman codeword, on the other hand, is
neither determined by the state's complexity nor by its probability.
Given $p_i$, the Huffman codeword length can, in principle, be as
small as 1~bit and as large as $[\log((\sqrt{5}+1)/2)]^{-1} \approx
1.44$ times $-\log p_i$ \cite{Katona1976}.

In this correspondence, we show that the lengths of both Huffman and
Shannon-Fano codewords have a similar interpretation. The probability
of the states for which the Huffman codeword length differs by more
than $m$~bits from $-\log p_i$ decreases exponentially with $m$. In
this sense, one can say that, for a {\it typical\/} state, the Huffman
codeword satisfies $l_i\approx-\log p_i$, just as for Shannon-Fano
coding.  This is especially relevant in a thermodynamic context where
entropies are of the order of $2^{80}$~bits and where an error of a
few hundred bits in the length of a typical record would be
unnoticeable.

\section*{Result}

In this section we return to the terminology of the abstract and
consider a discrete memoryless $N$-letter source ($N\geq2$) to which a
binary Huffman code is assigned. The $i$-th letter has probability
$p_i<1$ and codeword length $l_i$. The Huffman code can be represented
by a binary tree having the {\it sibling property\/}
\cite{Gallager1978} defined as follows: The number of links leading
from the root of the tree to a node is called the {\it level\/} of
that node.  If the level-$n$ node $a$ is connected to the
level-$(n+1)$ nodes $b$ and $c$, then $a$ is called the {\it parent\/}
of $b$ and $c$\/; $a$'s {\it children\/} $b$ and $c$ are called {\it
siblings}.  There are exactly $N$ terminal nodes or {\it leaves},
each leaf corresponding to a letter. Each link connecting two nodes is
labeled 0 or 1. The sequence of labels encountered on the path from
the root to a leaf is the codeword assigned to the corresponding
letter. The codeword length of a letter is thus equal to the level of
the corresponding leaf. Each node is assigned a probability such that
the probability of a leaf is equal to the probability of the
corresponding letter and the probability of each non-terminal node is
equal to the sum of the probabilities of its children. A tree has the
sibling property iff each node except the root has a sibling and the
nodes can be listed in order of nonincreasing probability with each
node being adjacent to its sibling in the list~\cite{Gallager1978}.

{\parindent=0pc \parskip=5pt

{\it Definition\/}: A level-$l$ node with probability $p$---or,
equivalently, a letter with probability $p$ and codeword length
$l$---has the property $X_m^+$ ($X_m^-$) iff $l>-\log p+m$ ($l<-\log
p-m$).

{\it Theorem 1\/}: $P_m^-=\sum_{j\in I_m^-}p_j<2^{-m}$ where
$I_m^-=\{i|l_i<-\log p_i-m\}$, i.~e., the  probability that a letter
has property $X_m^-$ is smaller than $2^{-m}$. (This
is true for any prefix-free code.)

{\it Proof\/}: $P_m^-=2^{-m}\sum_{j\in I_m^-}2^{\log p_j+m}<
2^{-m}\sum_{j\in I_m^-}2^{-l_j}\leq2^{-m}$. The last inequality
follows from the Kraft inequality.

{\it Lemma\/}: Any node with property $X_m^+$ has probability
$p<2^{-c(m-1)}$ where $c=(1-\log g)^{-1}-1\approx2.27$ with
$g=(\sqrt{5}+1)/2$.

{\it Proof\/}: Property $X_m^+$ implies $l>\lfloor-\log p+m\rfloor$
where $\lfloor x\rfloor$ denotes the largest integer less than or
equal to $x$. It is shown in Ref.~\cite{Katona1976} that, if $p$ and
$l$ are the probability and level of a given node, $p\geq1/F_n$
implies $l\leq n-2$ for $n\geq3$ where $F_n=[g^n-(-g)^{-n}]/\sqrt{5}
\geq g^{n-2}$ is the $n$-th Fibonacci number ($n\geq1$). Therefore, if
$\lfloor-\log p+m\rfloor\geq1$, the inequality $l>\lfloor-\log
p+m\rfloor$ implies $p<(F_{\lfloor-\log p+m\rfloor+2})^{-1}\leq
g^{-\lfloor-\log p+m\rfloor}\leq g^{\log p-m+1}$. For $\lfloor-\log
p+m\rfloor<1$, $p<g^{\log p-m+1}$ holds trivially.  Solving for $p$
proves the lemma.

{\it Theorem 2\/}: $P_m^+=\sum_{j\in I_m^+}p_j<2^{-c(m-2)+2}$ where
$I_m^+=\{i|l_i>-\log p_i+m\}$, i.~e., the  probability that a letter
has property $X_m^+$ is smaller than $2^{-c(m-2)+2}$.

{\it Proof\/}: Suppose there is at least one letter---and hence a
corresponding leaf---having the property $X_m^+$. Then, among all
nodes having the property $X_m^+$, there is a nonempty subset
with minimum level $n_0>0$. In this subset, there is a node having
maximum probability $p_0$. In other words, there is no node having
property $X_m^+$ on a level $n<n_0$, and on level $n_0$, there is no
node with probability $p>p_0$. Thus property $X_m^+$ implies
\begin{displaymath}
p_0 > 2^{-n_0+m} \;.
\end{displaymath}
Now let $k_0$ be the number of nodes on level $n_0-1$, and define the
integer $l_0<n_0$ such that $2^{l_0}\leq k_0<2^{l_0+1}$. Then the
number of level-$n_0$ nodes is less than $2^{l_0+2}$. Since all nodes
having property $X_m^+$ are on levels $n\geq n_0$, it follows that
\begin{displaymath}
P_m^+ < 2^{l_0+2} p_0 \;.
\end{displaymath}
In order to turn this into a useful bound, note the following.
The sibling property or, more directly, the optimality of a Huffman
code implies that all level-$(n_0-1)$ nodes have probability $p\geq
p_0$. Since there are at least $2^{l_0}$ level-$(n_0-1)$ nodes, it is
again a consequence of the sibling property that there exists a
level-$(n_0-1-l_0)$ node with probability $p_1 \geq 2^{l_0}p_0 >
2^{-n_0+m+l_0}$ and thus having property $X_{m-1}^+$. Using the
lemma, one finds $p_1 < 2^{-c(m-2)}$ and therefore
\begin{displaymath}
P_m^+ < 2^{l_0+2}p_0 \leq 2^2p_1 < 2^{-c(m-2)+2} \;.
\end{displaymath}

} 

\section*{Acknowledgements}

The author profited much from frank discussions with Carlton M.~Caves
and Christopher Fuchs.

\end{document}